# *In situ* twistronics of van der Waals heterostructures

Yaping Yang[1,2,*], Jidong Li[3], Jun Yin[3], Shuigang Xu[2], Ciaran Mullan[1], Takashi Taniguchi[4], Kenji Watanabe[4], Andre K. Geim[1,2], Konstantin S. Novoselov[1,2,5], Artem Mishchenko[1,2,*]

[1]School of Physics and Astronomy, University of Manchester, Oxford Road, Manchester, M13 9PL, UK

[2]National Graphene Institute, University of Manchester, Oxford Road, Manchester, M13 9PL, UK

[3]State Key Laboratory of Mechanics and Control of Mechanical Structures and MOE Key Laboratory for Intelligent Nano Materials and Devices, College of Aerospace Engineering, Nanjing University of Aeronautics and Astronautics, Nanjing, 210016, China

[4]National Institute for Materials Science, 1-1 Namiki, Tsukuba, 305-0044, Japan

[5]Centre for Advanced 2D Materials, National University of Singapore, 117546, Singapore

*e-mails: ypyang0916@gmail.com, artem.mishchenko@gmail.com

**In van der Waals heterostructures, electronic bands of two-dimensional (2D) materials, their nontrivial topology, and electron-electron interactions can be dramatically changed by a moiré pattern induced by twist angles between different layers. Such process is referred to as twistronics, where the tuning of twist angle can be realized through mechanical manipulation of 2D materials. Here we demonstrate an experimental technique that can achieve *in situ* dynamical rotation and manipulation of 2D materials in van der Waals heterostructures. Using this technique we fabricated heterostructures where graphene is perfectly aligned with both top and bottom encapsulating layers of hexagonal boron nitride. Our technique enables twisted 2D material systems in one single stack with dynamically tunable optical, mechanical, and electronic properties.**

Rotational misalignment caused by a twist angle between adjacent layers of 2D materials results in symmetry breaking and strain effects, leading to enhanced or suppressed interlayer coupling and electronic band structure reconstruction[1-4]. Moiré superlattices formed by rotational misalignment and lattice mismatch have been under intense scrutiny and an abundance of new phenomena have been observed[5-24]. Graphene placed on hexagonal boron nitride (hBN) represents a prototypical system of a twisted heterobilayer. In this twisted system, when at a small twist angle, mono- or few-layer graphene in a quantizing magnetic field exhibits a fractal energy spectrum known as Hofstadter's butterfly [5-7]. This twisted system has been reported to bear topological bands and strong correlations, exhibiting a number of fascinating many-body phenomena, such as correlated insulating states and superconductivity reported in trilayer graphene on hBN[8,9], and fractional Chern insulators observed in bilayer graphene on hBN[10]. In twisted homobilayers, strongly-correlated phenomena including correlated insulating states, unconventional superconductivity and ferromagnetism, have been observed in twisted bilayer graphene[11-14] and twisted bilayer-bilayer graphene systems[15-18]. A variety of other systems, such as twisted transition metal dichalcogenide layers[19-22], graphene/$WS_2$ bilayer[23], and twisted bilayer graphene aligned to hBN[24], have also manifested many intriguing phenomena.

In twistronics, accurate positioning, rotation, and manipulation of 2D materials are needed to fabricate a system with desired twist angles. To this end, several techniques have been developed, including optical alignment of crystal edges[7,25], tear and stack technique for twisted homobilayers[26], and *in situ* rotation mediated by atomic force microscope (AFM) tips[27]. Here we present a new experimental strategy to dynamically manipulate layered heterostructures *in situ* with precise control, allowing investigation of optical, mechanical, and electronic properties of a system with arbitrary twist angles between individual layers.

For our *in situ* twistronics technique, we use a glass slide with a droplet of polydimethylsiloxane (PDMS) as a manipulator, which is cured and naturally shaped into a hemisphere geometry. For a carefully fabricated PDMS hemisphere, the contact area between the manipulator and a 2D crystal can be as small as a few tens of micrometers[26,28], which depends on the hemisphere radius and is highly sensitive to the contact force, making it difficult to precisely control the target flake. To solve this problem, we developed a critical step, where we intentionally deposit an epitaxial polymethyl methacrylate (PMMA) patch on top of a target flake through a standard electron-beam lithography (EBL) process. The concept is illustrated in Fig. 1. The epitaxial PMMA patch could be designed into an arbitrary shape that fits the target flake and is normally a few hundred nanometers thick, thereby the contact area between the PDMS hemisphere and the flake is precisely limited to the area of the epitaxial PMMA patch without PDMS touching beyond the target flake at a much larger contact force threshold.

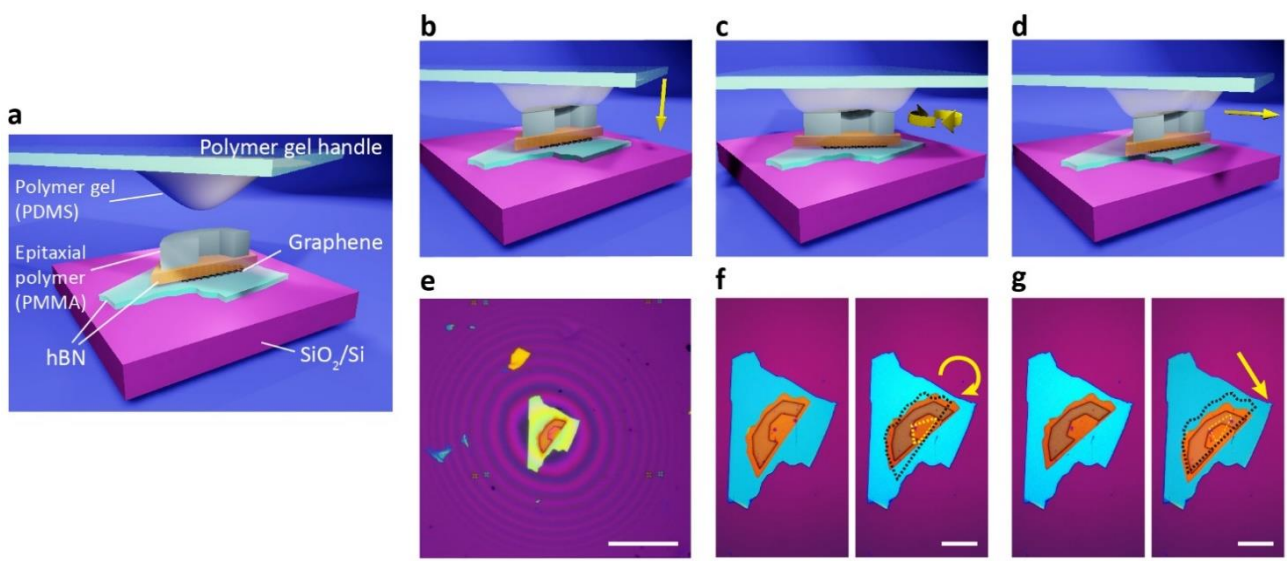

**Fig. 1: In situ manipulation of van der Waals heterostructures. a,** Schematic of polymer (PMMA)-mediated *in situ* manipulation technique. **b,** Schematic of the polymer gel handle (PDMS handle) touching PMMA. **c,** Schematic of rotating a 2D-material stack. **d,** Schematic of sliding a 2D-material stack. **e,** Optical image of the stack covered by a polymer (PMMA) patch in contact with the polymer gel (PDMS). The interference rings show the proximity of the PDMS hemisphere to the substrate. **f,** Optical images of the stack before (left) and after (right) rotation. The yellow arrow shows the rotation direction. **g,** Optical images of the stack before (left) and after (right) translation manipulation. The yellow arrow shows the translation direction. The dashed lines in the right panels of **f** and **g** indicate the original position of graphene and top hBN. The scale bars are 20 μm.

This strategy facilitates the accurate manipulation of the target flake. Figures 1b-d show the cartoon schematics of how the *in situ* manipulation of heterostructures works. To manipulate the target flake of the hBN/graphene/hBN heterostructure, first, by lowering down the polymer gel handle, PDMS hemisphere is brought in contact with the PMMA patch which is pre-patterned onto top hBN. When they touch, there is a colour change in the PMMA patch which can be easily distinguished under an optical microscope (see Fig. S1). Figure 1e shows the top view when PDMS hemisphere touches PMMA patch, where the interference rings imply that PDMS hemisphere is in the proximity of the flake without touching it. The friction between two incommensurately stacked 2D materials is dramatically reduced thanks to the superlubricity[29,30], and, therefore, the top hBN, as well as the underneath graphene, can not only slide but also rotate freely on the surface of the bottom hBN under the control of the PDMS hemisphere (Fig. 1c, d, f and g, and Supplementary Video 1 and 2). Figures 1f and g show that the PMMA patch is in good adhesion to the top hBN and does not delaminate during the manipulation process.

The manipulation technique presented here enables continuous modification of the twist angle between the layers even after the heterostructure is already assembled. In order to perform *in situ* optical measurements such as Raman spectroscopy after each manipulation step, the PMMA patch was intentionally designed not to cover the graphene so that it can give strong enough signal (Fig. 1f and g). Compared to previous *in situ* rotation technique mediated by an AFM tip[27], our manipulation technique is convenient and reproducible since the PMMA patch can be easily washed away by acetone and re-patterned by EBL. In addition, our technique can manipulate flakes regardless of their thickness, whereas an AFM tip might destroy thin flakes.

To demonstrate the potential of the manipulation technique in twistronics, we fabricated another hBN/graphene/hBN heterostructure (sample 1) where the graphene layer was aligned to both top and bottom hBN layers. We firstly assembled the heterostructure intentionally with $0 < \theta_t, \theta_b < \theta_{tb} < 60°$ (Fig. 2d), where $\theta_t$, $\theta_b$ and $\theta_{tb}$ are twist angles between graphene and top hBN, graphene and bottom hBN, as well as top hBN and bottom hBN, as shown in Fig. 2a and b. Then we used our technique to rotate the top hBN layer, by making sure that the rotating PMMA patch is patterned within the top layer (Fig. 2d). The twist angles $\theta_t$ and $\theta_b$ varied simultaneously as the rotation progressed until either of them reached 0° where the graphene layer was locked to one of the hBN layers, that is graphene/hBN interface went through a transition from incommensurate to commensurate state[31,32]. At a certain moment the smooth rotation stacks and the PMMA delaminated from the top hBN layer (Fig. 2e). We associate such conditions with $\theta_t$ and $\theta_b$ were both ≈0°, indicating that graphene was aligned to both hBN layers. At this stage, the two hBN layers were aligned to each other as well, with $\theta_{tb}$ also reaching 0°, as shown in Fig. 2e. The delamination of PMMA after all the 2D layers are locked to each other indicates that the interaction between the 2D materials at commensurate state is stronger than that between the PMMA and the top 2D crystal (see section 2 in SI).

In principle, there are two types of alignment in a doubly-aligned hBN/graphene/hBN heterostructure, that is $\theta_t=\theta_b=\theta_{tb}=0°$ and $\theta_t=0°$, $\theta_b=\theta_{tb}=60°$. The state of the resulting stack is determined by the initial settings of $\theta_t$ and $\theta_b$ and the rotation direction. Scenarios for various $\theta_t$ and $\theta_b$ are presented in Supplementary Fig. S2, where we also discuss in details the motion of incommensurately stacked 2D materials (section 2 in SI). The initial settings of the twist angles rely on the known crystal orientation of graphene and hBN layers, which is normally distinguished by straight long edges of the flake. Since hBN crystals are three-fold rotationally symmetric, using hBN flakes from the same exfoliated crystal can help to secure the crystal orientations. However, for hBN, the symmetry between the top and bottom atomic layers depends on whether the number of layers in hBN is odd or even (Fig. 2c). This ambiguity can, in principle, be circumvented by using the same surface, either top or bottom, of the original hBN crystal for alignment. To demonstrate how this can be done we fabricated sample 2 with $\theta_t=0°$, $\theta_b=\theta_{tb}=60°$ in the final stack (see section 2 and Figs. S3, S4 in SI).

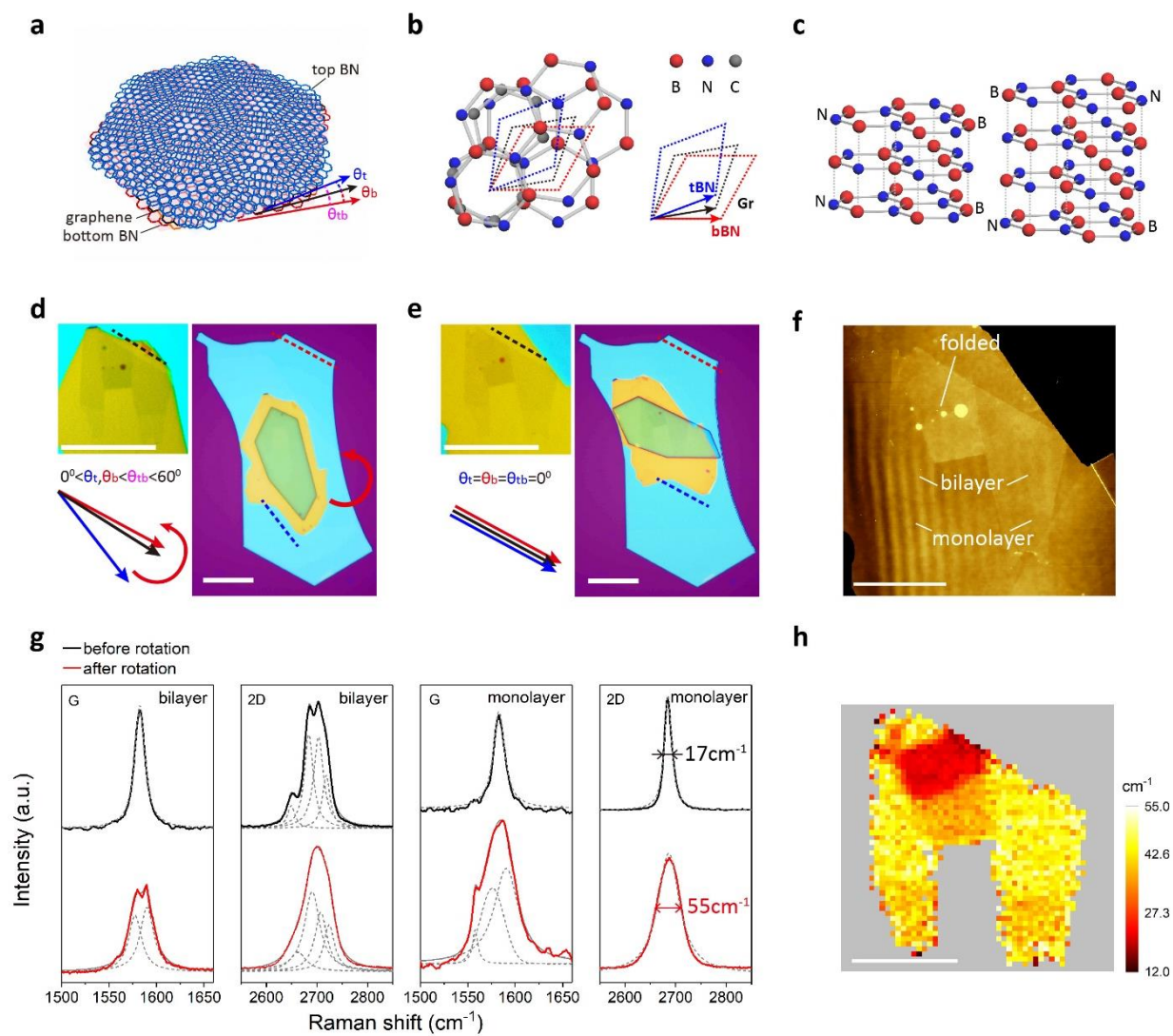

**Fig. 2: Encapsulated graphene perfectly aligned to both top and bottom hBN using *in situ* rotation technique.** **a**, Schematic of graphene encapsulated by hBN, with twist angles $\theta_t$, $\theta_b$ and $\theta_{tb}$ between the layers. **b**, Lattice structure of graphene encapsulated by hBN and the corresponding lattice vectors of each layer. **c**, Atomic structure of hBN, with odd (left) and even (right) numbers of layers. **d,e,** Optical images of the stack, before (d) and after (e) rotation. The top left panels increase the contrast to show the position of graphene. The bottom left panels show the relative crystal orientations of each layer. The red arc arrow highlights the rotation direction. The dashed lines indicate the crystal edges of graphene and hBN layers, which were aligned after the PMMA patch delaminated from the top hBN, as shown in **e**. The scale bars are 20 μm. **f,h**, AFM topography and 2D band width Raman map of the stack after rotation, respectively. The scale bars are 10 μm. **g**, Raman spectra of graphene at the monolayer and bilayer regions before and after rotation.

To confirm the alignment of graphene to both hBN layers, we carried out Raman characterization, as shown in Fig. 2g and h. The graphene layer in sample 1 originally contained monolayer and bilayer regions (Fig. 2f). For the monolayer region, the full width at half maximum of 2D peak (FWHM$_{2D}$) increases from 17 cm$^{-1}$ to 55 cm$^{-1}$ after rotation; G peak width also broadens with the emergence of lower frequency components (Fig. 2g). These results are consistent with previous reports[33,34]. The broadening of 2D peak by near 40 cm$^{-1}$ and the appearance of lower frequency components in G peak signify near-perfect alignment between the hBN and graphene crystals and arise from graphene coupling to the moiré potentials from both top and bottom hBN crystals[27,34]. For the bilayer region, the splitting of 2D peak is strongly enhanced when graphene is misaligned to both hBN layers, whereas after rotation, these components broaden and shift towards each other, resulting in a prominent change in the line shape of 2D peak. In contrast, G peak splits into two distinct components after rotation. These results for doubly-aligned bilayer graphene have not been reported before and we ascribe the change in Raman spectra to the periodic strain field induced by moiré patterns (see section 3 in SI). Figure 2h shows the 2D peak width Raman map of the graphene layer after rotation. It clearly demonstrates the homogeneous distribution of FWHM$_{2D}$ among different regions of the graphene layer, indicating a spatially uniform twist angle in the stack. The AFM topography (Fig. 2f) here shows that the rotation process did not damage or crease the graphene layer.

To further quantify the existence of two moiré superlattices at both sides of graphene layer in this heterostructure, we made it into a device and investigated its transport properties. We will now focus on our bilayer device, Fig. 3. Figure 3a shows the longitudinal resistivity $\rho_{xx}$ and transverse resistivity $\rho_{xy}$ as a function of charge carrier density $n$ at non-quantizing magnetic field of $B$ = 0.03 T. We observed both broadened resistivity peak at the primary Dirac point (PDP) and satellite resistivity peaks at finite densities situated symmetrically with respect to PDP. The Hall resistivity $\rho_{xy}$ near each satellite peak changes sign, suggesting that they are moiré-induced secondary Dirac points (SDPs).

In quantizing magnetic fields, the presence of moiré potential splits the Landau levels into a fractal structure: the Hofstadter minibands separated by a hierarchy of self-similar minigaps, to which the corresponding densities follow linear trajectories according to the Diophantine equation: $n/n_0 = t(\phi/\phi_0) + s$, where $s$ and $t$ are integers denoting the superlattice miniband filling index and quantized Hall conductivity of the gapped state, respectively, $\phi_0$ is the magnetic flux quantum and $n_0$ is the total number of electron states per area of a completely filled Bloch band[35]. Existence of two moiré patterns should result in two sets of such self-similar bands. Figure 3d is a simplified Wannier diagram showing the positions of the most prominent $\sigma_{xx}$ zeroes in the measured Landau fan diagram, Fig. 3b. We find that the Landau fans originate from PDP at $n$ = 0, and from the two sets of SDPs at $n_{s1}$ = ±2.15×10$^{12}$ cm$^{-2}$, and $n_{s2}$ = ±2.34×10$^{12}$ cm$^{-2}$ (corresponding to the moiré wavelengths of $\lambda_{s1}$ = 14.7 nm and $\lambda_{s2}$ = 14.0 nm, respectively). To identify the two SDPs more clearly, we plot $\partial\sigma/\partial n$ versus $B$ and $n$ in Fig. S8a, c and d in SI.

In the fractal superlattice spectra, the self-similar minigaps occur at $\phi = \phi_0 p/q$, where $\phi = BA$ is the magnetic flux per superlattice unit cell, $A$ is the superlattice unit cell area, $\phi_0$ is the magnetic flux quantum, $p$ and $q$ are co-prime integers. The strongest minigaps arise at $\phi = \phi_0/q$, resulting in Brown-Zak (BZ) magneto-oscillations[36] with the periodicity of $1/B = A/\phi_0$. These minigaps also correspond to the intersections between the central and satellite fans in the Landau fan diagram[5,6]. We observe Brown-Zak oscillations (horizontal streaks visible in Fig. 3b) belonging to the two sets of satellite fans, as shown by the horizontal blue and magenta dashed lines in Fig. 3d, which provides independent confirmation of the two distinct moiré periods. Brown-Zak oscillations are more clearly seen in $\partial\sigma/\partial B(B,n)$ maps in Fig. S8b, e and f, in which we observed other $p/q$ fractions. High temperature suppresses cyclotron oscillations, making the BZ oscillations more visible, Fig. S9. At electron doping where BZ oscillations are more pronounced[36], two sets of maxima, corresponding to the two periodicities, 15.2 nm and 14.7 nm, are clearly seen. This independently verifies the existence of two moiré superlattices with different wavelengths at both sides of the graphene layer.

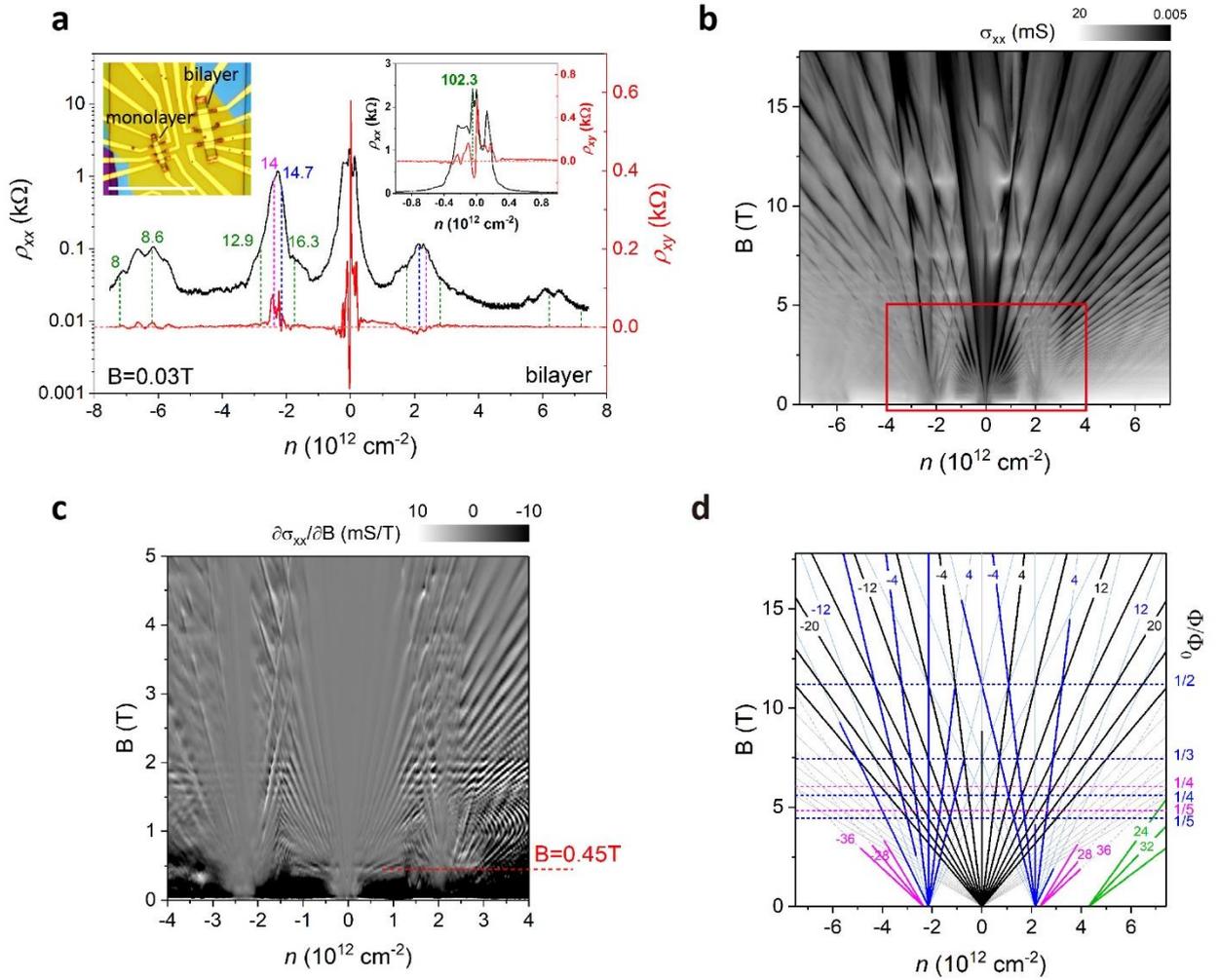

**Fig. 3: Hofstadter butterfly and Brown-Zak oscillations in bilayer graphene double moiré superlattices. a,** Longitudinal resistivity $\rho_{xx}$ and transverse resistivity $\rho_{xy}$ of the bilayer region as a function of charge carrier density $n$ after rotation. $T$ = 0.3 K, $B$ = 0.03 T. The dashed lines and numbers show the satellite peaks and the corresponding moiré wavelengths λ (see Methods). The blue and magenta dashed lines mark the SDPs at $n_{s1}$ and $n_{s2}$ corresponding to moiré patterns formed at both sides of graphene, respectively. The green dashed lines mark the $n_{sm}$ required to reach the first Brillouin zone edge of super-moiré pattern with different wavelengths. The left inset shows Hall bar devices made on monolayer and bilayer graphene regions. The scale bar is 20 μm. The right inset is the zoomed-in figure of **a**. **b,** Landau fan diagram $\sigma_{xx}(n,B)$ measured at $T$ = 0.3 K. **c,** Fan diagram $\partial\sigma_{xx}/\partial B(n,B)$ highlighting the Brown-Zak oscillations in part (red rectangular) of **b**. The BZ feature at $B$ = 0.45 T

originates from the first order magnetic Bloch state of the super-moiré pattern with the largest $\lambda_{sm} \approx 102.3$ nm at $\phi = \phi_0$. **d**, Simplified Wannier diagram labelling the quantum Hall states identified in **b**. Quantum oscillations with dominant sequence of Landau level filling factors $v = \pm4, \pm8, \pm12, ...$ emerge from the PDP (black lines). Blue lines show quantum oscillations of $t = \pm4, \pm8, \pm12, ...$ emerging from SDP $n_{s1}$. Magenta lines show quantum oscillations of $t = \pm20, \pm28, \pm36$ emerging from SDP $n_{s2}$. Green lines show another set of gap trajectories ($s = 2$) from the same moiré superlattice as SDP $n_{s1}$ according to the Diophantine equation, with quantum oscillations of $t = 24, 32$ and $44$. The blue and magenta horizontal dashed lines and numbers on the right show the most prominent BZ oscillations belonging to the SDPs $ns_1$ and $ns_2$, respectively, with different values of p/q for $\phi = (p/q) \phi_0$.

Such coexisting moiré patterns with different wavelengths should in principle interfere, resulting in a second-order (composite) moiré pattern, as reported in experimental[37-39] and theoretical studies[40,41]. From the two SDPs we calculated the corresponding twist angles to be 0.24° and 0.38° (see Methods). Using the method described in Refs.[37,38], we obtained six possible composite moiré wavelengths ($\lambda_{sm}$, see section 4 in SI), and observed satellite peaks in $\rho_{xx}$ near most of the corresponding carrier densities $n_{sm}$ required to reach the first Brillouin zone edge of the six possible composite moiré patterns, which are marked by green dashed lines and numbers in Fig. 3a. The $n_{sm}$ of the largest $\lambda_{sm} \approx 102.3$ nm is in the range of $\pm0.04\times10^{12}$ cm$^{-2}$ to $\pm0.05\times10^{12}$ cm$^{-2}$ (right inset of Fig. 3a). Notably, in the plot of $\partial\sigma_{xx}/\partial B(n,B)$ within $|n| < \pm2\times10^{12}$ cm$^{-2}$ we observe a prominent horizontal streak at $B = 0.45$ T (Fig. 3c), which perfectly matches the magnetic field of the first-order magnetic Bloch state originating from the composite moiré pattern when $BA = \phi_0$ (where $q = 1$). The results for the bilayer graphene region discussed above are similar to what was found in the monolayer graphene region (Figs. S10-S12). The manipulation technique presented here proves successful in making the hBN/graphene/hBN heterostructures with perfect alignment between graphene and the two hBN layers with a high twist angle homogeneity. In contrast, optical alignment of crystal edges or heating of the final stack cannot consistently guarantee perfect alignment with the twist angle < 1° in a single stack[5-7,10,37,39].

In conclusion, we introduce an *in situ* manipulation of van der Waals heterostructures mediated by patterning a polymer resist patch onto target flakes, which can precisely and dynamically control the rotation and positioning of 2D materials by a simple gel stamp manipulator. Using this technique, we realized a perfect alignment of graphene with respect to hBN layers with a much higher success rate compared to conventional optical alignment of crystal edges during micromechanical transfer. Our technique, which can be easily generalized to other 2D material systems (as shown in Fig. S5), opens up a new strategy in device engineering for twistronics, finding its applications in the design of 2D quasicrystals[42,43], magic-angles flat bands[11-14], devices with AB/BA domain walls[44] and other topologically nontrivial systems.

**Acknowledgements**
A.M. acknowledges the support of EPSRC Early Career Fellowship EP/N007131/1.

## Methods

### 1. Van der Waals assembly

All heterostructures were assembled using standard dry-transfer technique[45] using a polymethyl methacrylate (PMMA) coated on a polydimethyl siloxane (PDMS) stamp, and a SiO$_2$ (290 nm)/Si as a substrate. The devices made from the monolayer and bilayer graphene regions are in a Hall bar geometry, with electrical contacts made by Cr/Au (3 nm/80 nm).

### 2. Raman characterization

The Raman spectra were acquired by Renishaw Raman system with 1800 lines/mm grating, using linearly polarized laser radiation at the wavelength of 532 nm. The laser power was kept below 5 mW. The Raman spatial maps were taken with a step size of 0.5 μm. To extract the peak width of 2D band, we fit the spectrum at each pixel in the spatial mapping to a single Lorentzian function.

### 3. Transport measurements

Transport measurements were carried out in a four-terminal geometry with a low-frequency ac current excitation of 100 nA using standard lock-in technique at the base temperature of 0.3 K (Brown-Zak oscillations were measured at 70 K). The devices were gated to control a charge carrier density ($n$) and a displacement field ($D$) by applying bias voltages to the metal top gate ($V_t$) and the doped silicon substrate ($V_b$). The charge carrier density is determined by $n = (D_b - D_t)/e$, the vertical displacement field $D$ is set by $D = (D_b - D_t)/2$. Here $D_b = \varepsilon_b(V_b - V_b^0)/d_b$, and $D_t = -\varepsilon_b(V_t - V_t^0)/d_t$, where $\varepsilon_{t,b}$ and $d_{t,b}$ are the dielectric constants and thicknesses of the top and bottom dielectric layers, respectively, and $V_b^0$ and $V_t^0$ are the effective offset voltages caused by environment-induced doping.

### 4. Calculation of moiré wavelength, twist angle and super-moiré wavelength

The moiré wavelength $\lambda$ is calculated using the geometric relation $n_s = 4/\frac{\sqrt{3}}{2}\lambda^2$, where $n_s$ is the charge density at full filling of the moiré miniband (at the density of four electrons per superlattice unit cell). The relation between twist angle $\theta$ and moiré superlattice wavelength $\lambda$ in hBN/graphene/hBN heterostructure is given by[46]:

$$\lambda = \frac{(1+\delta)a}{\sqrt{2(1+\delta)(1-\cos\theta)+\delta^2}} ,$$

$$\tan\phi = \frac{-\sin\theta}{(1+\delta)-\cos\theta}$$

where $a$ is the graphene lattice constant, $\delta$ is the lattice mismatch between hBN and graphene which is 1.65% (we used the value calculated by Ref [33,38]), $\phi$ is the relative orientation of the two moiré patterns with respect to the graphene lattice ($\phi_1$ and $\phi_2$, respectively).

Using the method described previously[37], for super-moiré wavelength $\Lambda$, in analogy to the equation above, the relation between $\Lambda$ and the twist angle $\Theta$ between the constituent moiré patterns (the wavelengths being $\lambda_1$ and $\lambda_2$, $\lambda_1 \geq \lambda_2$) is given by:

$$\Lambda = \frac{(1+\Delta)\lambda_2}{\sqrt{2(1+\Delta)(1-\cos\Theta)+\Delta^2}}$$

where $\Delta$ is the mismatch between the constituent moiré patterns, given by:

$$\Delta = \frac{\lambda_1 - \lambda_2}{\lambda_2}$$

and the twist angle $\Theta$ is given by:

$$\Theta = (|\phi_1 - \phi_2| - 30°)\,mod(60°) - 30°$$

The modular division used here to find $\Theta$ reproduces the output of the piecewise conditional reported in Ref 37.

# Supplementary information

**1. Identification of the initial contact between polymer gel and epitaxial polymer**

In our manipulation technique, the critical step to precisely control the movement of 2D flakes is the identification of the initial contact between the polymer gel (PDMS) and the epitaxial polymer patch (PMMA). The signature of the contact is a colour change in the PMMA patch and an appearance of a sharpen edge of the contact area between the PDMS gel and PMMA patch, which is easily distinguished under optical microscope, Fig. S1.

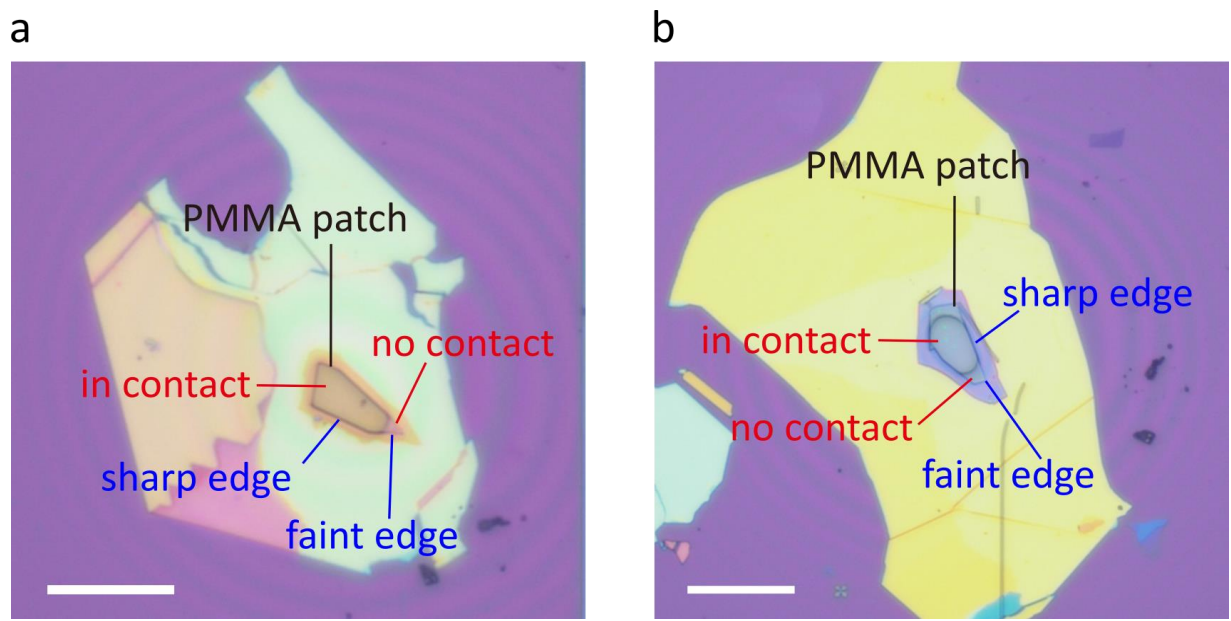

**Fig. S1.** Micrographs showing the moment when polymer gel (PDMS) is brought into contact with epitaxial polymer (PMMA). (a) PDMS gel is in contact with most of the PMMA patch, only in a small region on the right corner the contact is absent. The region in contact shows a different colour from the no-contact region. (b) PDMS gel touches the centre of the PMMA patch, with the in-contact area limited to the PMMA patch. In both images, the edges of the in-contact regions are thicker than the edges of no-contact regions. The scale bars are 50 μm.

**2. The motion of incommensurately stacked 2D materials**

In our manipulation technique, the motion of 2D materials in the van der Waals heterostructure depends on the balance between external driving force and the adhesion between PMMA patch and the top 2D layer, as well as the kinetic friction originating from the atomic shear force between the adjacent 2D layers. The kinetic friction between 2D layers is dramatically reduced in the incommensurate state, the so called superlubricity where the 2D flakes can move smoothly. Inversely, in the commensurate state, the two adjacent 2D layers are locked together due to pinning of the boundaries that separate local regions of the commensurate phase[1-3]. Such boundaries, either topological defects or misfit dislocations, are caused by the lattice mismatch and twist angle between adjacent layers.

For graphene/hBN and hBN/hBN interfaces, the commensurate state occurs at a small twist angle between the layers, since these two interfaces already satisfy the condition of a small lattice mismatch (1.65%[4,5] for graphene/hBN interface and 0% for hBN/hBN interface). The locking of the graphene/hBN and hBN/hBN interfaces limits further motion of the flakes at commensurate state, thereby when the external driving force is large enough, the PMMA patch/hBN/graphene/hBN stack will break at the weaker coupled PMMA patch/hBN interface, in agreement with the delamination of PMMA patch when graphene is aligned to both hBN layers after rotation (Fig. 2e in the main text and Fig. S3c). Rotation of the top two layers of hBN/graphene/hBN heterostructure with different initial settings of $\theta_t$ and $\theta_b$ should result in the final stack with commensurate or incommensurate states at the two interfaces, as shown in Fig. S2 and Table S1. Here we consider both the 0° and 60° alignment of graphene and hBN layer, since these two types of alignment have distinct symmetry and affect electronic structure of graphene differently[4,6].

We fabricated hBN/graphene/hBN heterostructure (sample 2) with the manipulation process matching the cases in Fig. S2b (initial alignment) and Fig. S2j (final alignment). To make sure that the crystal orientation of the flakes is known, we first identified hBN fakes which have fractured into two pieces during the mechanical exfoliation procedure (Fig. S3a). Note that for hBN crystals with odd numbers of layers, the top and bottom layers have the same crystal orientation, whereas for hBN crystals with even numbers of layers, the top and bottom layers are of mirror symmetry (Fig. 2c in the main text). Therefore, the common edges of the fractured hBN pieces indicate the same crystal orientation of the atomic layer belonging to the same surface, either top or bottom, of the original crystal. Then we used one of the fractured hBN piece to pick up graphene and flipped over the other fractured piece as the bottom hBN layer, so that the hBN atomic layers adjacent to graphene come from the same crystal surface. See Fig. S4 for the details of this procedure. Next, the initial $\theta_t$ and $\theta_b$ were set according to Fig. S2b and the following rotation direction led to the final stack with $\theta_t=0°$, $\theta_b=\theta_{tb}=60°$ (Fig. S3c), which is the case of Fig. S2j.

We carried out Raman spectroscopy to prove the alignment of graphene and hBN in sample 2, as shown in Fig. S3d and e. The behaviour of Raman spectra in monolayer and bilayer graphene regions are similar to those observed in sample 1. Whereas the full width at half maximum of 2D peak ($FWHM_{2D}$) increased from 17 cm$^{-1}$ to 48.5 cm$^{-1}$ after rotation, smaller than that in the perfect alignment case, which means that the twist angles $\theta_t$ and $\theta_b$ are larger than those in sample 1. Figure S3e shows the distribution of $FWHM_{2D}$ among the graphene flake after rotation, indicating a spatially uniform twist angle in both monolayer and bilayer regions. The AFM topography (Fig. S3f) here shows that the rotation process did not damage or crease the graphene layer.

The assembly process of sample 2 is illustrated in Fig. S4. The details of the process are as follows. **Step 1**, we used PMMA coated PDMS block mounted on a glass slide to pick-up one of the fractured hBN pieces as the top hBN (hBN piece 1). **Step 2**, we used polypropylene carbonate (PPC) coated PDMS block mounted on another glass slide to pick-up the other fractured hBN piece (hBN piece 2), which serves as the bottom hBN. **Step 3**, the target graphene flake is picked up by PDMS/PMMA/hBN piece 1. Now the graphene flake is in contact with the bottom surface of hBN piece 1. **Step 4**, the PDMS/PPC block is flipped over so that the bottom surface of hBN piece 2 faces upwards, as indicated by the red line in the figure. Then the PDMS/PMMA/ hBN piece 2 is rotated in order to align the shared crystal edges of hBN piece 1 and 2 under microscope. **Step 5**, hBN piece 2 is picked up by PMMA/hBN piece 1 /graphene at temperature above 75°. When temperature is high enough, the adhesion between PPC and hBN is smaller than the adhesion between PMMA and hBN. Thereby the bottom hBN (hBN piece 2) can be easily picked up by PMMA/hBN piece 1 /graphene. **Step 6**, the final stack hBN piece 1/graphene/ hBN piece 2 is released onto a new silicon wafer by dissolving away PMMA using acetone. In the final stack, graphene flake is in contact with both the bottom surfaces of hBN piece 1 and hBN piece 2. Since the bottom surfaces of hBN piece 1 and hBN piece 2 are originally from the same atomic layer, we can guarantee the settings of $\theta_t$ and $\theta_b$ if the alignment of the shared edges is based on the same surface of the fractured hBN pieces.

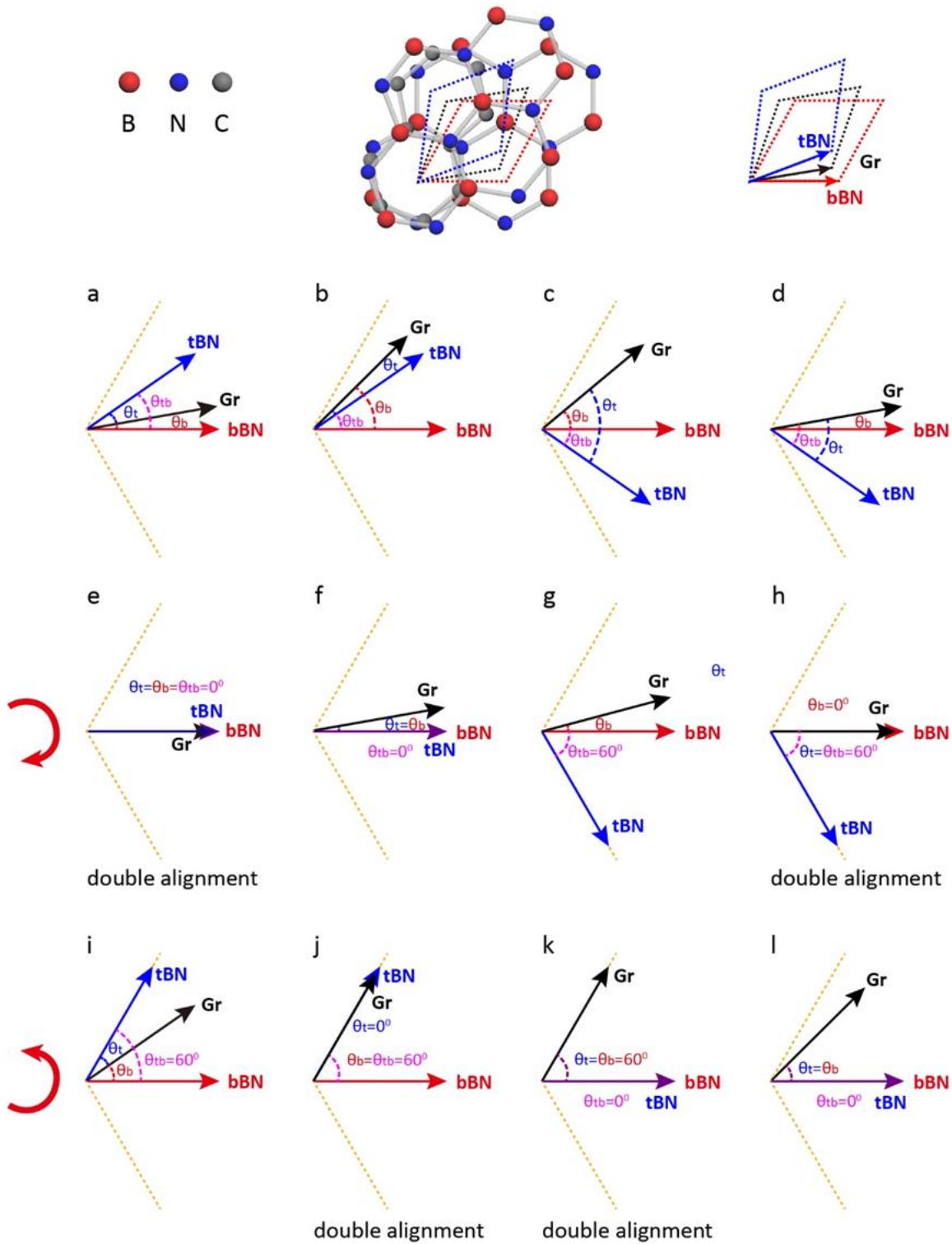

**Fig. S2.** The resulting alignment of hBN/graphene/hBN stack under different rotation directions. (a) to (d) hBN/graphene/hBN stack with different initial settings of $\theta_t$ and $\theta_b$. (e) to (h) The final stacks corresponding to (a) to (d) with different types of alignment after clockwise rotating. (i) to (l) The final stacks corresponding to (a) to (d) with different types of alignment after anti-clockwise rotating. The red, black and blue arrows indicate the crystal orientation of bottom hBN, graphene and top hBN, respectively. The red arc arrows on the left show the rotation direction. During the rotation process, the bottom hBN layer remains at original position and only the top hBN and graphene rotate under the control of polymer gel manipulator. Orange dashed lines mark ±60° with respect to red arrows (bottom hBN), the rotations will stack at these lines as top hBN goes into AA' stacking with bottom hBN.

**Table S1.** Resulting stack with different initial settings of $\theta_t$ and $\theta_b$ as shown in Fig. S2.

| Panels in Fig. S2 | Initial setting of twist angles (°) | Resulting stack | |
|---|---|---|---|
| | | Rotation direction | |
| | | clockwise | anticlockwise |
| a | $0<\theta_t, \theta_b<\theta_{tb}<60°$ | $\theta_t=0°, \theta_b=0°, \theta_{tb}=0°$, double alignment | $\theta_t\neq 0°, \theta_b\neq 0°, \theta_{tb}=60°$, misalignment |
| b | $0<\theta_t, \theta_{tb}<\theta_b<60°$ | $\theta_t=\theta_b\neq 0°, \theta_{tb}=60°$, misalignment | $\theta_t=0°, \theta_b=\theta_{tb}=60°$, double alignment |
| c | $-60°<\theta_{tb}<0<\theta_b<60°, |60°-\theta_{tb}|<\theta_b$ | $\theta_t=\theta_b\neq 0°, \theta_{tb}=60°$, misalignment | $\theta_t=\theta_b=60°, \theta_{tb}=0°$, double alignment |
| d | $-60°<\theta_{tb}<0<\theta_b<60°, |60°-\theta_{tb}|>\theta_b$ | $\theta_b=0°, \theta_t=\theta_{tb}=60°$, double alignment | $\theta_t=\theta_b\neq 0°, \theta_{tb}=0°$, misalignment |

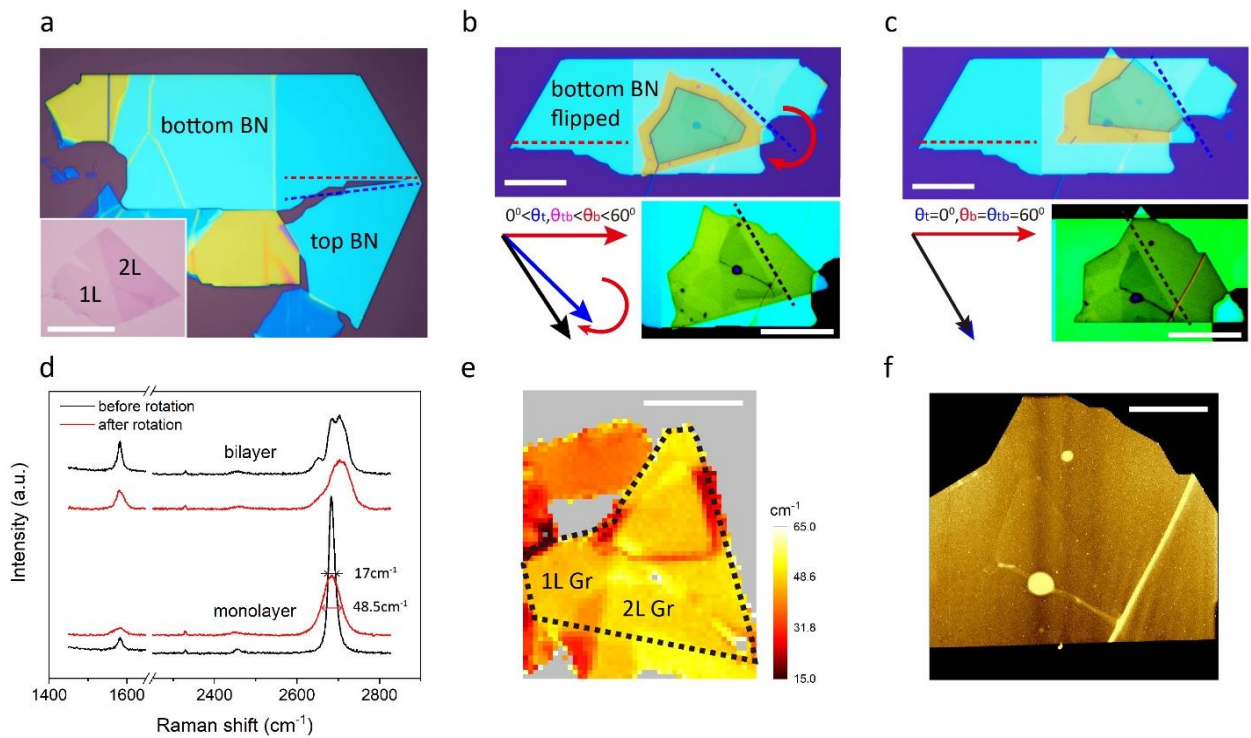

**Fig. S3.** Encapsulated graphene aligned to both top and bottom hBN, with $\theta_t=0°$, $\theta_b=\theta_{tb}=60°$. (a) Optical image of an hBN flake with a crack after mechanical exfoliation. The bottom inset shows the graphene flake containing monolayer and bilayer region. (b) Bottom hBN was flipped over before the release of top hBN/graphene stack. The final heterostructure was intentionally designed with the relative crystal orientations $0<\theta_t$, $\theta_{tb}<\theta_b<60°$, as shown in left bottom inset, which matches the case in Fig. S2b. The red arc arrow implies the rotation direction. (c) The heterostructure after rotation, with $\theta_t=0°$, $\theta_b=\theta_{tb}=60°$ as shown in left bottom inset, which matches the case in Fig. S2j. The PMMA patch delaminated from top hBN after all the crystals were aligned to each other. The right bottom insets of (b) and (c) increase the contrast to show the position of graphene. The red, blue and black dashed lines in (a-c) indicate the crystal orientation of bottom hBN, top hBN and graphene respectively. (d) Raman spectra of graphene at the monolayer and bilayer region before and after rotation, respectively. (e) 2D band width Raman map of the stack after rotation. The dashed line shows the region where graphene is covered by top hBN. (f) AFM topography of the stack after rotation. The scale bars in (a) to (c) are 20 μm. The scale bars in (e) and (f) are 10 μm.

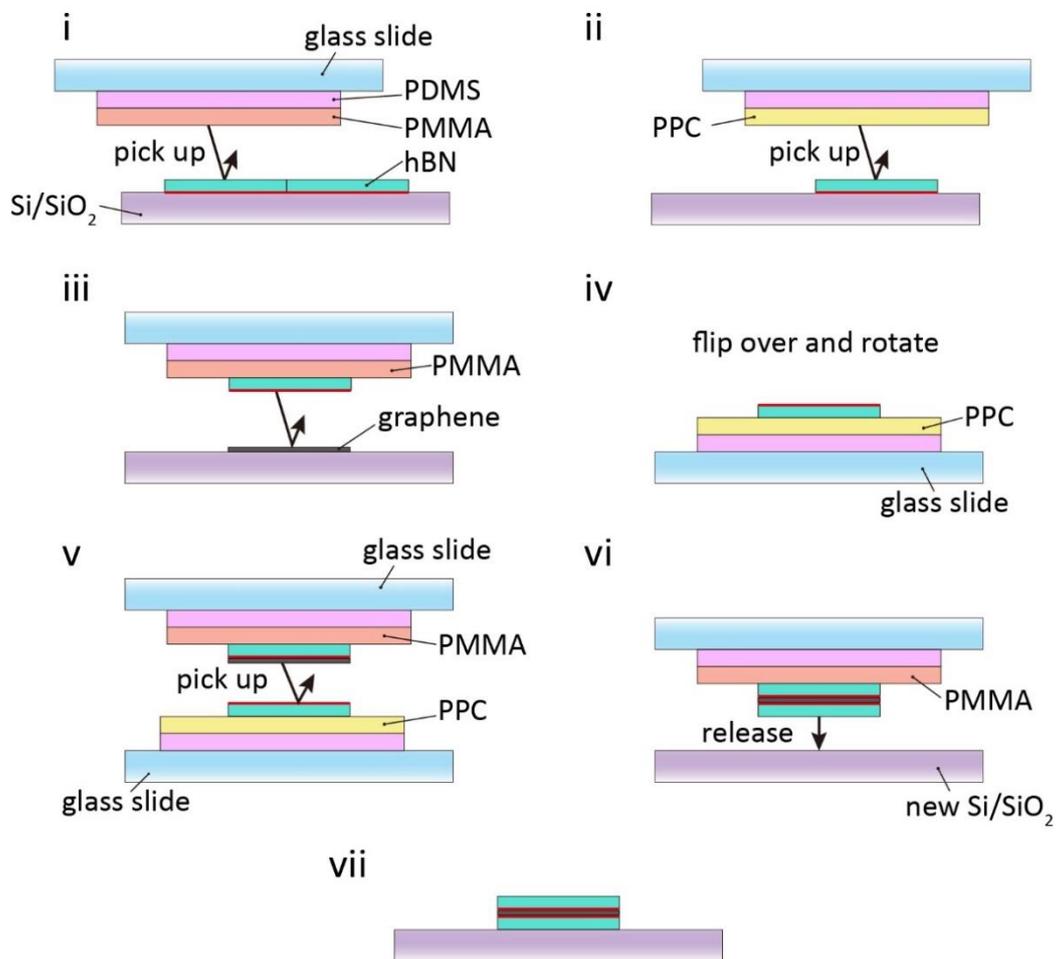

**Fig. S4.** Schematic process flow for graphene encapsulated by the same hBN crystal. In the final stack, the hBN atomic layers adjacent to graphene belong to the same surface of the original crystal, as highlighted by the red lines.

For other interfaces where the adjacent layers have larger lattice mismatch, such as $MoS_2$/hBN interface, a small or even zero twist angle will not result in the commensurate state. The kinetic friction between the 2D layers always remains at an extremely low value, thereby the 2D flake can move and rotate smoothly even when passing through the point where twist angle is zero without the delamination of PMMA patch, as shown in Fig. S5.

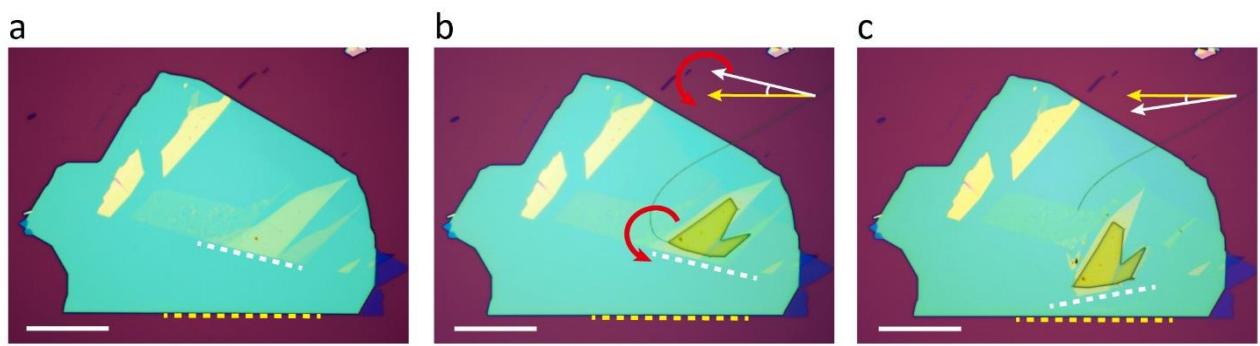

**Fig. S5.** Rotation of MoS$_2$ on top of hBN. (a) Optical image of MoS$_2$/hBN stack before rotation. The white and yellow dashed lines indicate the crystal orientations of MoS$_2$ and hBN crystal, respectively. (b) PMMA patch patterned onto MoS$_2$ flake. The white and yellow arrows show the twist angle between MoS$_2$ and hBN crystal. The red arc arrow implies the rotation direction. (c) Rotation of MoS$_2$ flake with the twist angle passing through 0°, without PMMA being delaminated from MoS$_2$ flake, confirming that there is no commensurate state at a small twist angle of MoS$_2$ and hBN crystal. The scale bars are 20 μm.

## 3. Raman spectra of graphene in the monolayer and bilayer region before and after rotation

Raman spectra of graphene are strongly modified when graphene is subjected to double moiré superlattices (Fig. 2g). In our sample, for monolayer graphene region, G peak changes its shape from a narrow peak centred at 1582 cm$^{-1}$ with FWHM$_G$ ≈14.5 cm$^{-1}$ to a broadened asymmetric peak after rotation. The presence of a low energy shoulder at ≈1558 cm$^{-1}$ of G peak after rotation is attributed to a TO phonon, similar to what was observed in graphene aligned to only one hBN crystal[7]. For 2D peak, the broadening effect is strongly enhanced in double-moiré superlattice compared to a simple aligned graphene/hBN heterostructure: ≈20 cm$^{-1}$ increase in FWHM for graphene in a single moiré superlattice[7,8] and ≈40 cm$^{-1}$ increase in FWHM for our doubly-aligned sample, indicating that the double moiré superlattices induce a much stronger periodic inhomogeneity originated from charge accumulation, strain, etc.

For bilayer graphene region, G peak splits into two components in the presence of double moiré superlattices. For 2D peak, the four components broadened and their position differences reduced, which is similar to what was reported in bilayer graphene aligned to only one hBN crystal[9].

We found an overall downshift of around 2cm$^{-1}$ for both G peak and 2D peak among the whole flake after rotation (Fig. S6 a-d), which is contradictory to the results in previous study[4,8]. Similar behaviour is found in Sample 2 (Fig. S7). Near the corner of the folded region (as indicated by the arrows), before rotation, the peak positions of G and 2D peaks are slightly lower than those of the nearby region. We attribute this phonon softening to the strain caused by folding[10]. Whereas after rotation, peak position near the corner of the folded region shows an upshift behaviour, and is higher than that of the nearby region. Note that during the rotation, the folded graphene rotated as well, therefore we expect an enhanced strain effect near this region. However the abnormal behaviour of the peak position implies a mechanism beyond strain effect.

To further look into the line shape of G peak after rotation, we fit G peak with two Lorentzian peaks for both bilayer and monolayer graphene regions and plot the maps of the peak position difference and intensity ratio of the two components, as shown in Fig. S6e and f. The maps show that the G peak splitting is homogeneous for both bilayer and monolayer graphene regions, which is around 13 cm$^{-1}$ and 22 cm$^{-1}$, respectively. The intensity ratio of the two components is around 1 for the bilayer region, which means that they have comparable weight, and is around 0.6 for the monolayer region.

The overall downshift and broadening of the peak positions for G and 2D peaks and splitting of G peak remind one of the effects of strain. Under uniaxial strain, the doubly degenerate $E_{2g}$ mode splits into two components, one along the strain direction and the other perpendicular, which leads to the splitting of G peak[11]. Tensile strain usually gives phonon softening, while compressive strain usually leads to phonon stiffening. In the presence of moiré superlattice, the strain spatial distribution is periodically modulated, and tension and compression of graphene lattice should coexist. The combined effect of the opposite types of strain could explain the observed behaviour of the Raman peaks.

Alternatively, the shift of G and 2D peaks could be related to a change in the Kohn anomalies associated with the phonon dispersion at Γ and K. For instance, the aligned hBN layers in the hBN/graphene/hBN heterostructure could have built-in electric fields[12] which can dope graphene and thereby shift the Raman peak positions. However, this scenario is unlikely, as previous studies show that the presence of electrostatic doping causes non-adiabatic removal of the Kohn anomaly from the Γ point, leading to the stiffening of G band and 2D band[13,14]. Thus, most likely, the main factor that causes the change in Raman spectra is the periodic strain field induced by moiré patterns.

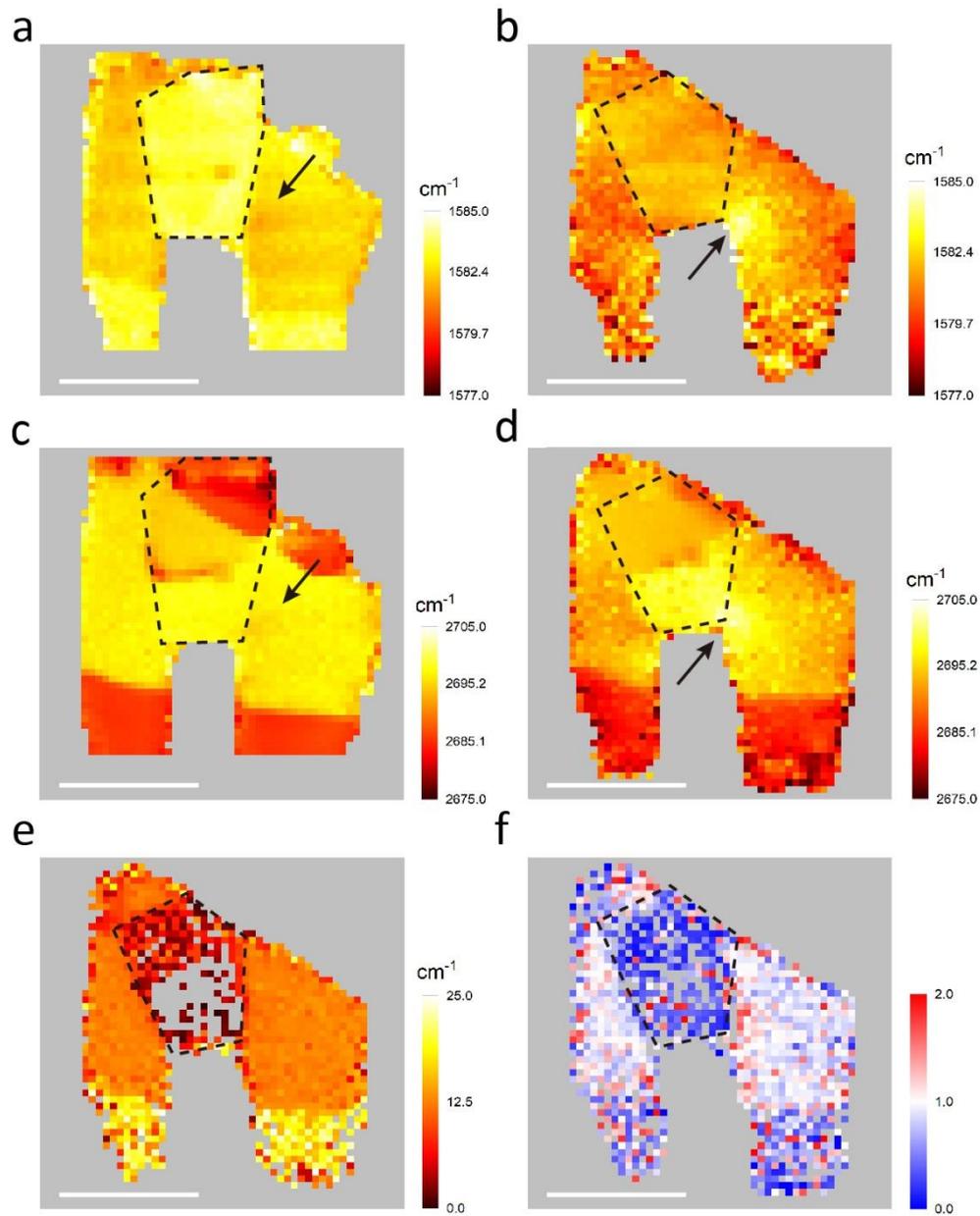

**Fig. S6.** Raman maps of sample 1 before and after rotation. (a) G peak position map before rotation. (b) G peak position map after rotation. (c) 2D peak position map before rotation. (d) 2D peak position map after rotation. G and 2D peaks of (a) to (d) were fitted by a single Lorentzian. (e) Map of the position difference of the two components of G peak after rotation. (f) Map of the peak intensity ratio of G peak components after rotation. G peak of (e) and (f) was fitted by two Lorentzian peaks. The scale bars are 10 μm. The black dashed lines show the folded region. The black arrows indicate a higher strain level near the folded corner.

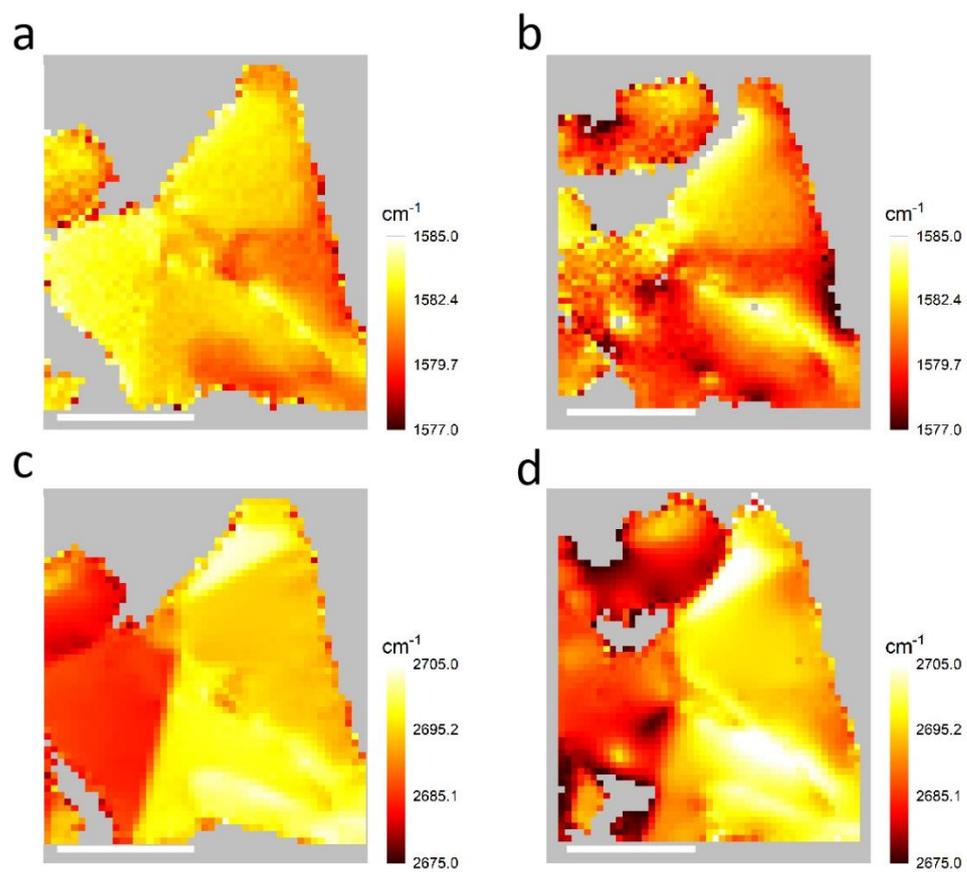

**Fig. S7.** Raman maps of sample 2 before and after rotation. (a) G peak position map before rotation. (b) G peak position map after rotation. (c) 2D peak position map before rotation. (d) 2D peak position map after rotation. G and 2D peaks were fitted by a single Lorentzian. The scale bars are 10 μm.

**4. Transport properties of graphene in double moiré superlattices.**

In commensurate magnetic field and periodic moiré potential, the fractal energy spectrum of electron in graphene exhibits a self-similar recursive, where the superlattice minibands show a gapped Dirac-like spectrum at $\phi = \phi_0 p/q$ and exhibit Landau levels (LLs) characteristic of Dirac fermions[15]. Therefore we expect to observe quantized Hall conductivity associated with each minigap, similar to what is observed in usual quantum Hall effect where $\sigma_{xy} = \nu e^2/h$, $\nu$ is a filling factor of a particular LL. Accordingly, when $\sigma_{xx}$ is plotted in a map vs $n$ and $B$ (fan diagram), the gaps between Landau levels (the zeroes in the $\sigma_{xx}$) should trace linear trajectories emerging from primary Dirac point (PDP) as well as secondary Dirac points (SDPs) induced by moiré superlattice. For the fans emerging from PDP, the trajectories follows $B = n\phi_0/\nu$. For the fans emerging from SDPs, substituting $\phi$ with $BA$ and $n_0=1/A$ in the Diophantine equation $\phi/\phi_0 = (n/n_0-s)/t$ yields $B-(-s/t)\phi_0/A = n\phi_0/t$ (which can be viewed as the quantization of minibands at an effective magnetic field of $B_{eff} = B-(-s/t)\phi_0/A$ or $B_{eff} = B-(p/q)\phi_0/A$, where $p$ and $q$ are co-prime integers). Thus we can deduce the corresponding carrier density to SDP from fan diagram.

Based on the Raman spectra of the hBN/graphene/hBN heterostructure, we estimate that the twist angles $\theta_t$ and $\theta_b$ should be both close to 0°, thereby in the transport properties, we expect to see two sets of SDPs close to each other apart from the PDP. For both bilayer and monolayer graphene devices, in longitudinal resistivity ($\rho_{xx}$) vs carrier density ($n$) dependence, we observed broadened (and with a complex structure) satellite peaks located around $n = \pm 2\times 10^{12}$ cm$^{-2}$ (Fig. 3a in the main text and Fig. S10a). At small $B = 0.03$ T where the Landau quantisation is not yet developed, the transversal resistivity $\rho_{xy}$ changes sign in the carrier density regions of these satellite peaks, indicating that they are moiré superlattice induced SDPs. These SDPs are more prominent in the hole side compared to the electron side, which is consistent with previous studies[16,17].

In the Landau fan diagram at high magnetic fields, by tracing the linear trajectories we observed two sets of SDPs induced by the two independent moiré patterns for both bilayer and monolayer device. For bilayer device, the SDPs are at $n_{s1} = \pm 2.15\times 10^{12}$ cm$^{-2}$, and $n_{s2} = \pm 2.34\times 10^{12}$ cm$^{-2}$ (Fig. 3a, b and d in the main text and Fig. S8a, c and d), corresponding to the moiré wavelengths of $\lambda_{s1} = 14.7$ nm and $\lambda_{s2} = 14.0$ nm, and the twist angles of 0.24° and 0.38°, respectively. For monolayer device, the SDPs are at $n_{s1} = \pm 2.10\times 10^{12}$ cm$^{-2}$, $n_{s2} = \pm 2.44\times 10^{12}$ cm$^{-2}$ (Fig. S10 and Fig. S11a, c and d), corresponding to the moiré wavelengths of $\lambda_{s1} = 14.8$ nm and $\lambda_{s2} = 13.8$ nm, and the twist angles of 0.2° and 0.43°, respectively. In addition to the Raman map in Fig. 2h in main text, the similar values of the two sets of SDPs obtained from the bilayer and the monolayer devices, confirm that top and bottom twist angles are spatially uniform in the stack.

In the fan diagram $\partial\sigma/\partial B(n,B)$, we observed Brown-Zak (BZ) oscillations originating from the first order magnetic Bloch states ($\phi/\phi_0=1/q$) for the two sets of SDPs in both bilayer and monolayer devices (Fig. S8b, e, f and Fig. S11b, e, f). The first order magnetic Bloch states formed by the two moiré superlattices are also prominent at high temperature ($T = 70$ K, where the Landau quantisation is suppressed), as indicated by the arrows in Figs. S9 and S12. We also observed high order magnetic Bloch states ($\phi/\phi_0 = 3/q$) belonging to the two moiré superlattices, as shown in Fig. S8e, f and Fig. S11e, f.

In principle, the super-moiré pattern generated by the two original moiré patterns should have six possible reciprocal lattice vectors, as described in the previous study[5], whereas the method in Ref[18] only considers one possible reciprocal lattice vector with the largest super-moiré pattern wavelength ($\lambda_{sm}$ = 102.3 nm). If we take into account other possible reciprocal lattice vectors, we will get six possible second-order moiré wavelengths in total. For bilayer graphene device, the other five super-moiré wavelengths are 16.3 nm, 12.9 nm, 8.6 nm, 8 nm and 7.2 nm. The corresponding carrier densities $n_{sm}$ of the first Brillouin zone edge are $\pm 1.75 \times 10^{12}$ cm$^{-2}$, $\pm 2.8 \times 10^{12}$ cm$^{-2}$, $\pm 6.2 \times 10^{12}$ cm$^{-2}$, $\pm 7.2 \times 10^{12}$ cm$^{-2}$ and $\pm 9 \times 10^{12}$ cm$^{-2}$. For monolayer graphene device, the six super-moiré wavelengths are 64.7 nm, 17.8 nm, 12.1 nm, 8.9 nm, 7.8 nm and 7.2 nm. The corresponding carrier densities $n_{sm}$ of the first Brillouin zone edge are $\pm 0.11 \times 10^{12}$ cm$^{-2}$, $\pm 1.46 \times 10^{12}$ cm$^{-2}$, $\pm 3.14 \times 10^{12}$ cm$^{-2}$, $\pm 5.8 \times 10^{12}$ cm$^{-2}$, $\pm 7.5 \times 10^{12}$ cm$^{-2}$ and $\pm 8.9 \times 10^{12}$ cm$^{-2}$. We observed satellite peaks in $\rho_{xx}$ near most of these carrier densities $n_{sm}$ (Fig. 3a in the main text and Fig. S10a). Note that near $n_{sm}$ corresponding to these resistivity peaks, most of the low-$B$ $\rho_{xy}$ regions have sign reversal. In addition, similar to bilayer device, in the plot of $\partial\sigma_{xx}/\partial B(n,B)$ of monolayer device, we observe prominent horizontal streaks at $B$ = 1.14 T and 0.57 T, which perfectly matches the magnetic field of the first and second order magnetic Bloch state originating from the super-moiré pattern when $BA = \phi_0$ (where $q$ = 1) and $BA = \phi_0/2$ (where $q$ = 2), respectively. These features allow us to attribute the origin of the observed carrier densities $n_{sm}$ to the super-moiré pattern with different wavelengths.

## 4.1 Bilayer graphene in double moiré superlattices.

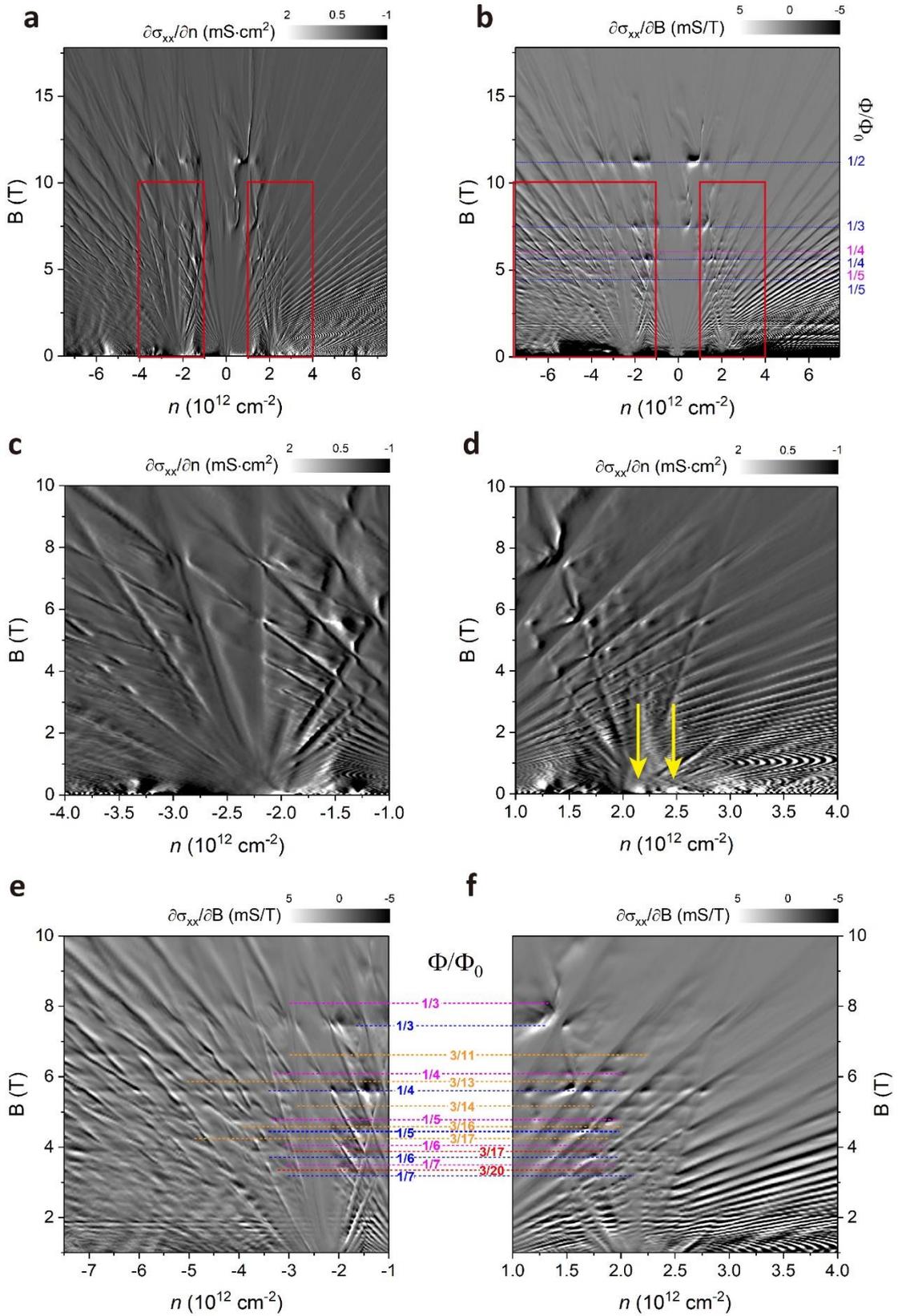

**Fig. S8.** Fractal Landau fan diagrams of bilayer graphene in double moiré superlattices. (a) Fan diagram $\partial\sigma_{xx}/\partial n(n,B)$ at $T$ = 0.3 K. (b) Fan diagram $\partial\sigma_{xx}/\partial B(n,B)$. (c) and (d) are part of (a) for hole and electron doping (marked in (a) by red rectangles), respectively, near secondary Dirac points (SDPs). Yellow arrows indicate the two sets of SDPs. (e) and (f) are part of (b) for hole and electron doping (marked in (b) by red rectangles),

respectively, near SDPs. The blue and red dashed lines show the Brown-Zak oscillations belonging to the SDP at $n_{s1} = \pm 2.14 \times 10^{12}$ cm$^{-2}$, whereas magenta and orange dashed lines belong to the SDP at $n_{s2} = \pm 2.34 \times 10^{12}$ cm$^{-2}$. The numbers on the dashed lines show the values of $p/q$ for oscillations at $\phi = (p/q)\phi_0$.

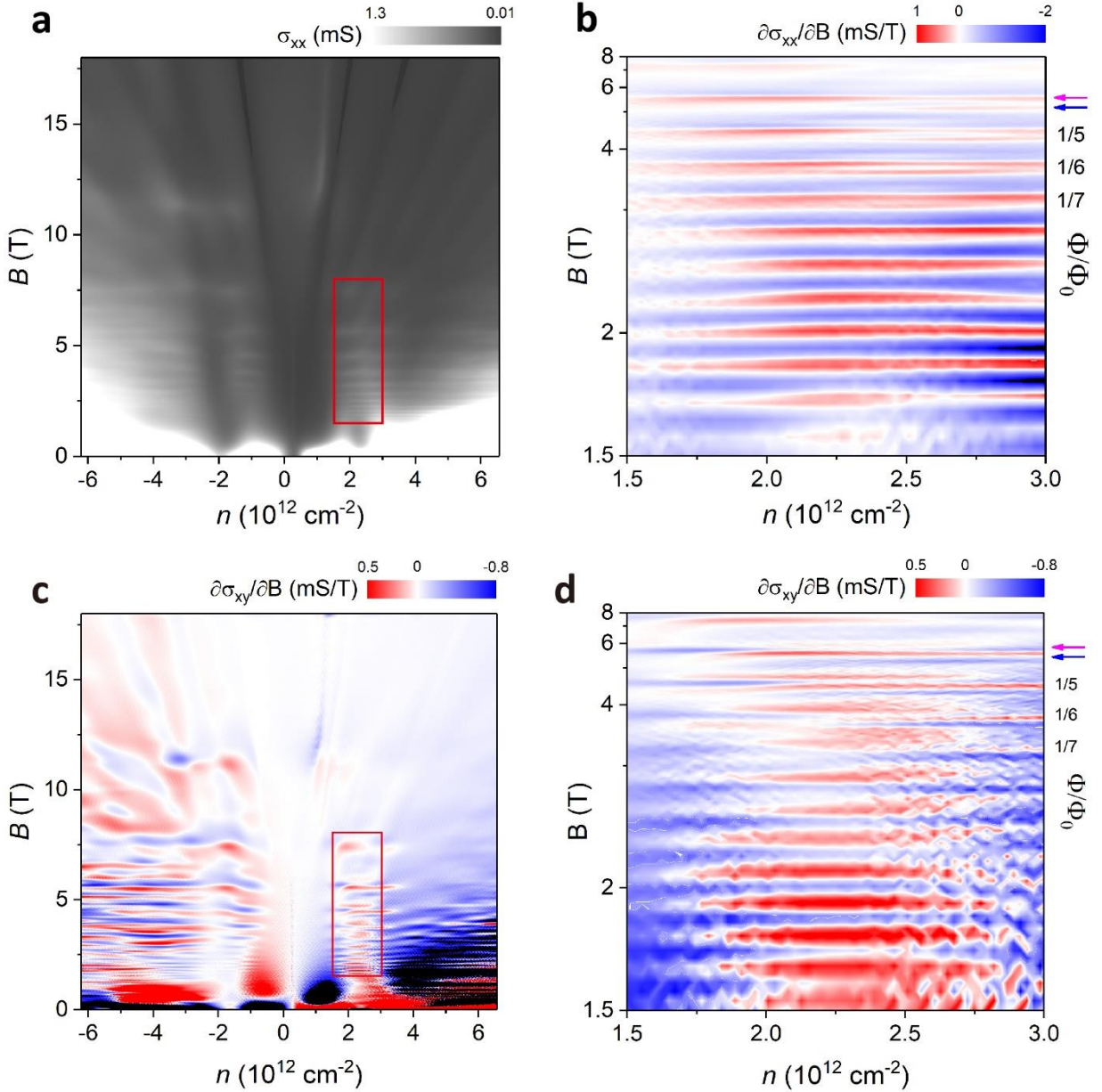

**Fig. S9.** Brown-Zak oscillations in double moiré bilayer graphene at $T$ = 70 K. (a) Longitudinal conductivity $\sigma_{xx}$ as a function of $n$ and $B$, $\sigma_{xx}(n,B)$. (b) $\partial\sigma_{xx}/\partial B(n,B)$ for part of (a) near the SDPs for electron doping (marked in (a) by a red rectangle). (c) $\partial\sigma_{xy}/\partial B(n,B)$. (d) Region of (c) near the SDPs for electron doping (marked in (c) by a red rectangle). The blue and magenta arrows show the Brown-Zak oscillations belonging to the SDPs at $n_{s1} = \pm 2.14 \times 10^{12}$ cm$^{-2}$, and $n_{s2} = \pm 2.34 \times 10^{12}$ cm$^{-2}$, respectively. The numbers on the right in (b) and (d) indicate $\phi = (1/q)\phi_0$ with $q$ = 5 to 7. At low magnetic fields, the resolution in $B$ is not sufficient to distinguish the two sets of oscillations.

*4.2 Monolayer graphene in double moiré superlattices.*

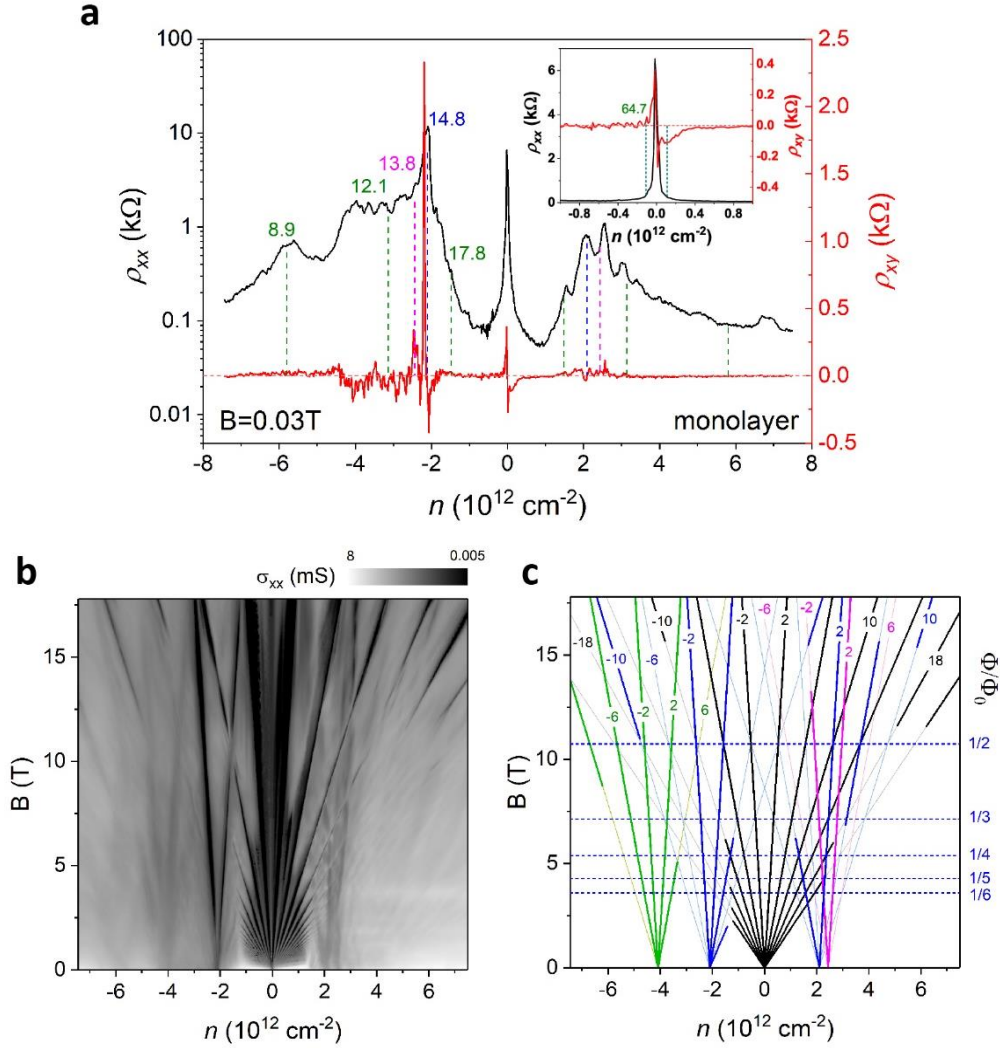

**Fig. S10.** Electronic transport and Landau quantization of monolayer graphene in double moiré superlattices. **a**, Longitudinal resistivity $\rho_{xx}$ and transverse resistivity $\rho_{xy}$ of the monolayer region as a function of charge carrier density $n$ after rotation. $T$ = 0.3 K, $B$ = 0.03 T. The dashed lines and numbers show the satellite peaks and the corresponding moiré wavelengths λ (see Methods). The blue and magenta dashed lines mark the SDPs at $n_{s1}$ and $n_{s2}$ corresponding to moiré patterns formed at both sides of graphene, respectively. The green dashed lines mark the $n_{sm}$ required to reach the first Brillouin zone edge of super-moiré pattern with different wavelengths. The right inset is the zoomed-in figure of **a**. **b**, Fan diagram $\sigma_{xx}(n,B)$ measured at $T$ = 0.3 K. **c**, Simplified Wannier diagram labelling the quantum hall effect states identified in **b**. Black lines show quantum oscillations with dominant sequence of Landau level filling factor $v$ = ±2, ±6, ±10, ... emerging from the PDP. Blue lines show quantum oscillations of $t$ = ±2, ±6, ±10, emerging from SDP $n_{s1}$. Magenta lines show quantum oscillations of $t$ = ±2 emerging from SDP $n_{s2}$. Green lines are another set of gap trajectories ($s$ = 2) from the same moiré superlattice as SDP $n_{s1}$ according to the Diophantine equation, with quantum oscillations of $t$ = ±2, ±6. The blue horizontal dashed lines and numbers on the right show the most prominent BZ oscillations belonging to the SDP $n_{s1}$, with different values of p/q for ϕ = (p/q) ϕ$_0$.

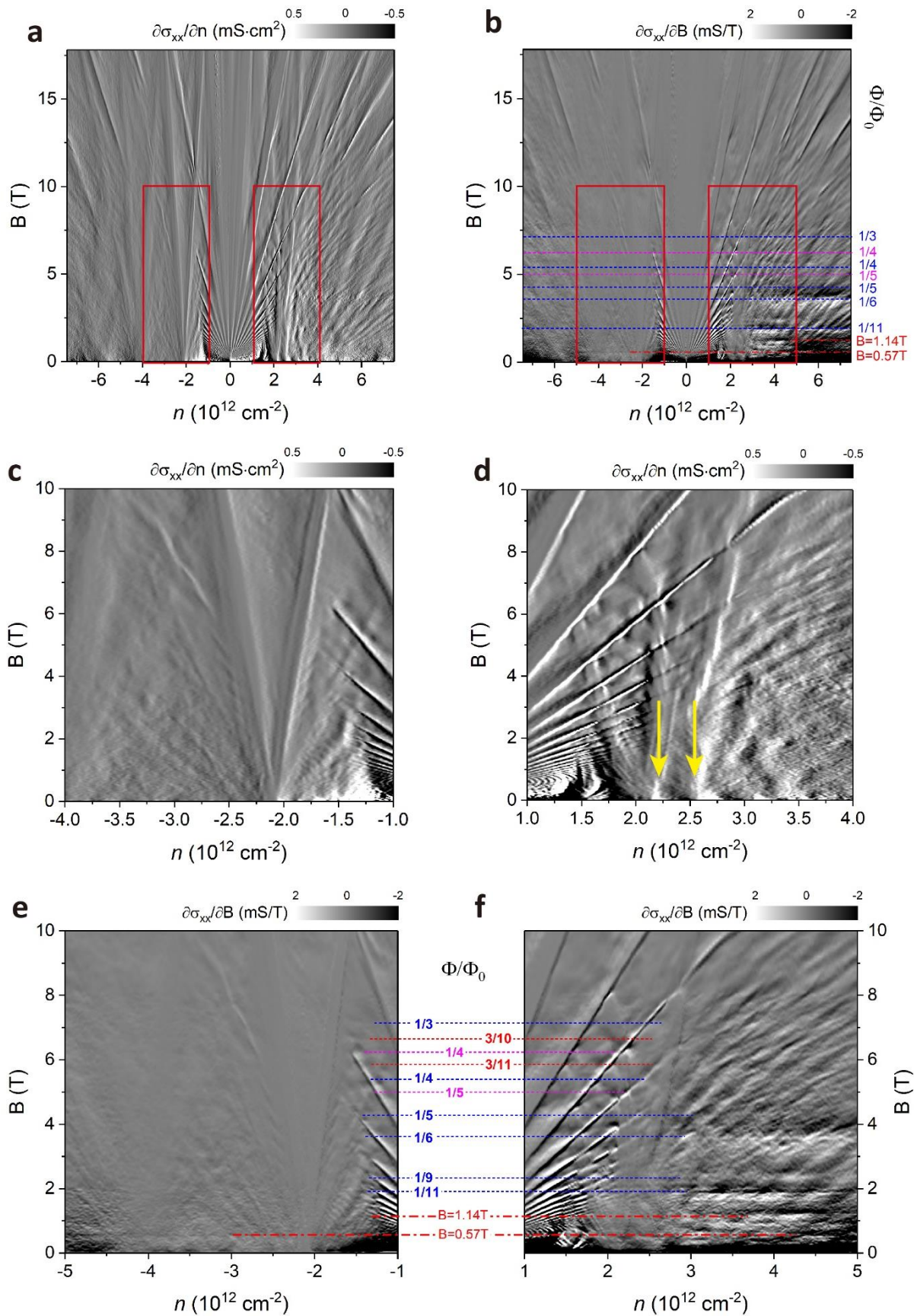

**Fig. S11.** Fractal Landau fan diagrams of monolayer graphene in double moiré superlattices. (a) Fan diagram $\partial\sigma_{xx}/\partial n(n,B)$ at $T$ = 0.3 K. (b) Fan diagram $\partial\sigma_{xx}/\partial B(n,B)$. (c) and (d) are part of (a) for hole and electron doping

near SDPs (marked in (a) by red rectangles), respectively. Yellow arrows indicate the two sets of SDPs. (e) and (f) are part of (b) for hole and electron doping (marked in (b) by red rectangles), respectively, near SDPs. The blue and red dashed lines show the BZ oscillations belonging to the SDPs at $n_{s1} = \pm 2.10 \times 10^{12}$ cm$^{-2}$. The magenta dashed lines show the BZ oscillations belonging to the SDP $n_{s2} = \pm 2.44 \times 10^{12}$ cm$^{-2}$. The numbers on the dashed lines show the values of $p/q$ for oscillations at $\phi = (p/q)\phi_0$. The BZ feature at $B = 1.14$ T and 0.57 T originates from the first and second order magnetic Bloch state of the super-moiré pattern with the largest $\lambda_{sm} \approx 64.7$ nm at $\phi = \phi_0$ and $\phi = \phi_0/2$, respectively.

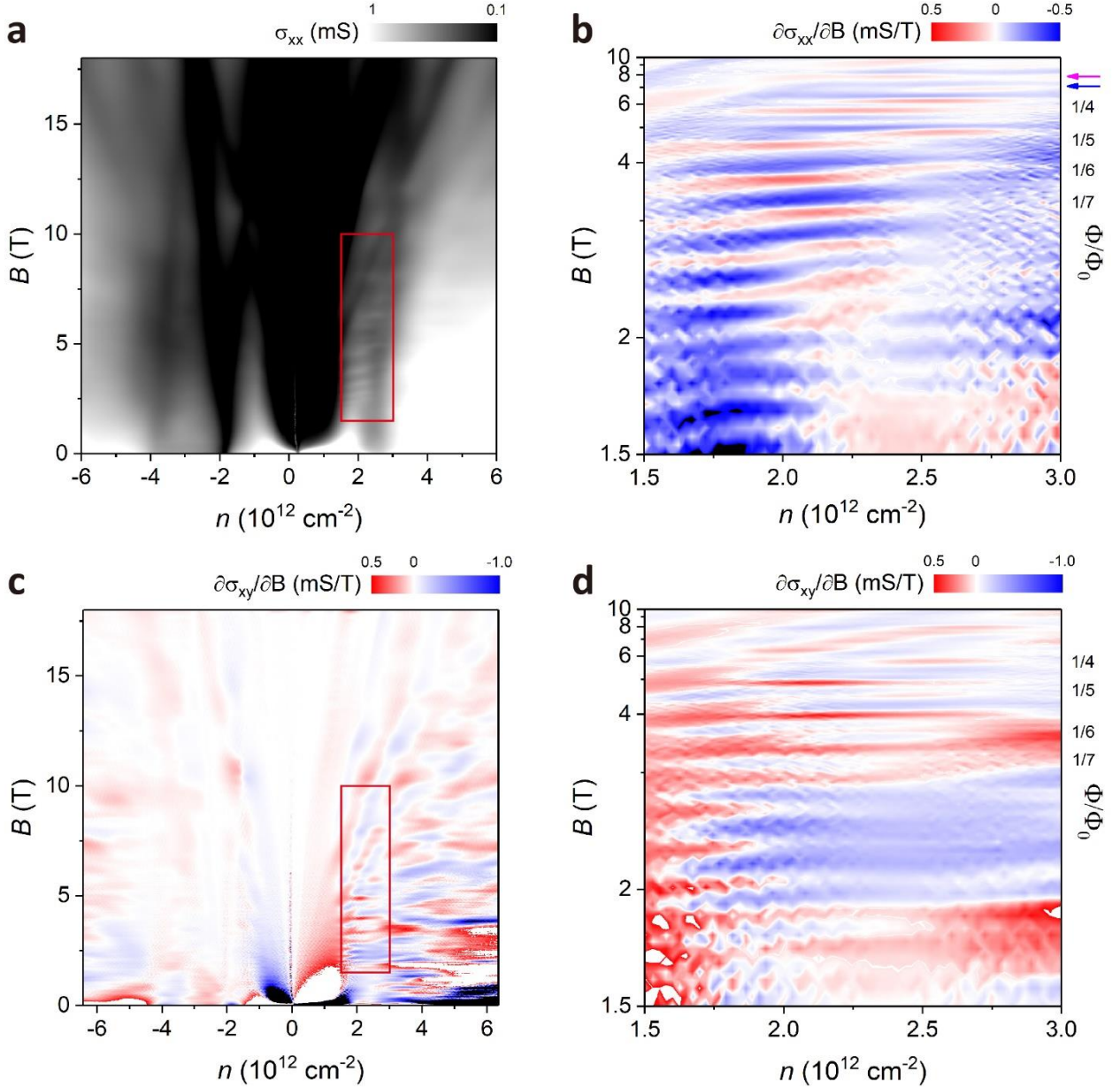

**Fig. S12.** Brown-Zak oscillations in double moiré monolayer graphene at $T = 70$ K. (a) Longitudinal conductivity $\sigma_{xx}$ as a function of $n$ and $B$, $\sigma_{xx}(n,B)$. (b) $\partial\sigma_{xx}/\partial B(n,B)$ in part of (a) near the SDP for electron doping (marked in (a) by a red rectangle). The blue and magenta arrows show the Brown-Zak oscillations belonging to the SDPs at $n_{s1} = \pm 2.10 \times 10^{12}$ cm$^{-2}$, and $n_{s2} = \pm 2.44 \times 10^{12}$ cm$^{-2}$, respectively. (c) $\partial\sigma_{xy}/\partial B(n,B)$. (d) Part of (c) near the SDP for electron doping (marked in (c) by a red rectangle). The numbers on the right in (b) and (d) indicate $\phi = (1/q)\phi_0$ with $q = 4$ to 7.